\documentclass{article}

\usepackage[final,nonatbib]{nips_2016}

\usepackage[utf8]{inputenc} 
\usepackage[T1]{fontenc}    
\usepackage{hyperref}       
\usepackage{url}            
\usepackage{booktabs}       
\usepackage{amsfonts}       
\usepackage{nicefrac}       
\usepackage{microtype}      
\usepackage{xcolor}         
\usepackage{mathtools}

\usepackage{footnote}
\usepackage{cleveref}

\usepackage{graphicx} 
\usepackage{subfigure}

\crefformat{footnote}{#2\footnotemark[#1]#3}

\newcommand{\Comments}{1}
\newcommand{\note}[2]{\ifnum\Comments=1\textcolor{#1}{#2}\fi}

\newcommand*\samethanks[1][\value{footnote}]{\footnotemark[#1]}

\title{Hemingway: Modeling Distributed Optimization Algorithms}

\author{Xinghao Pan\thanks{Joint first authors},~ Shivaram Venkataraman\samethanks[1],~ Zizheng Tai,~ Joseph Gonzalez \\
  \{xinghao, shivaram, zizheng.tai, jegonzal\}@cs.berkeley.edu \\
  UC Berkeley
}

\begin{document}

\maketitle


\begin{abstract}
Distributed optimization algorithms are widely used in many industrial machine learning
applications. 
However choosing the appropriate algorithm and cluster size is often difficult for users
as the performance and convergence rate of optimization algorithms vary with the size of the cluster.
In this
paper we make the case for an ML-optimizer that can select the appropriate algorithm and cluster size to use
for a given problem. To do this we propose building two models: one that captures the system level
characteristics of how computation, communication change as we increase cluster sizes and another
that captures how convergence rates change with cluster sizes. We present preliminary results
from our prototype implementation called Hemingway and discuss some of the challenges involved in
developing such a system.
\end{abstract} 


\section{Introduction}
With growing data sizes and the advent of cloud computing infrastructure~\cite{venkataraman2016ernest},
distributed machine learning is used in a number of applications like machine translation, computer vision, speech recognition etc.
As a result, recent research has proposed a number of distributed optimization algorithms~\cite{zhang15b,jaggi14,duchi2011adaptive} that handle large input datasets~\cite{bottou10} and minimize communication~\cite{yang13,li14b,jaggi14} to scale across large clusters.

The performance (or time to converge to $\epsilon$ error) of distributed optimization algorithms depends on the cluster setup used for training.
For example, assuming a data-parallel setup, the
time taken for one iteration of full gradient descent depends on the time taken to \emph{compute} the gradient in parallel and the time
taken to \emph{communicate} the gradient values. As the cluster size increases the computation time decreases but the communication time increases and thus choosing an optimal cluster size is important for optimal performance~\cite{venkataraman2016ernest}.

However, in addition to performance, the convergence rates of algorithms
also change based on the size of the cluster used. For example, in
CoCoA~\cite{jaggi14}, a communication efficient dual coordinate ascent algorithm,
each machine executes a local learning procedure and the resulting dual vectors are then averaged across
machines at the end of each iteration. 
With a fixed data size, using a larger number of machines will
lower the time spent in local learning but lead to a worse convergence rate. 
Thus, as we increase
the cluster size, the time per-iteration decreases but the number of iterations required to reach the desired error increases.

Further, as the computation vs. communication balance and convergence rates differ across algorithms 
(e.g., first-order methods~\cite{bottou10} vs. second-order methods~\cite{qu2015sdna}), it is often
hard to predict which algorithm will be the most appropriate for a given cluster setup. Finally, the 
convergence rates also depend on data properties. Thus while theoretical analysis can provide upper 
bounds on the number of iterations required, it is often difficult to translate this to how the 
algorithm will work in practice for a given dataset.

The combination of the above factors in conjunction with the dependence on the data, complicates the choice of optimization algorithm and cluster configuration.  
Choosing the best algorithm, cluster setup thus becomes a trial-and-error process and users have few tools
that can guide them in this process.

In this paper, we propose addressing this problem by building a system that can 
model the convergence rate of various algorithms and thereby help users select the best one for
their use case. To do this we propose modeling the convergence
rate of algorithms as we scale the cluster size and we split our
modeling into two parts: based on Ernest~\cite{venkataraman2016ernest}, we
first build a computational model that helps us understand how the time taken
per-iteration varies as we increase the scale of computation; we then build a
separate model for the convergence rate as we scale the computation and we show
how combining these two models can help us determine the optimal configuration.

We propose Hemingway, a prototype implementation and present initial evaluation results from running our system 
with CoCoA+~\cite{ma15}. We also outline some of the remaining challenges in making such a system
practical. These include designing data acquisition methods that can minimize the amount of data required to build
the above mentioned models and also extending our work to non-convex domains like deep-learning.


\section{Background}

\begin{figure*}[t!]
  \centering
  \subfigure[Time per iteration as we vary the degree of parallelism used. The plot shows the mean across 50 iterations and the error bars show the 5th and 95th percentile]{\label{fig:time-per-iter}\includegraphics[width=0.30\textwidth]{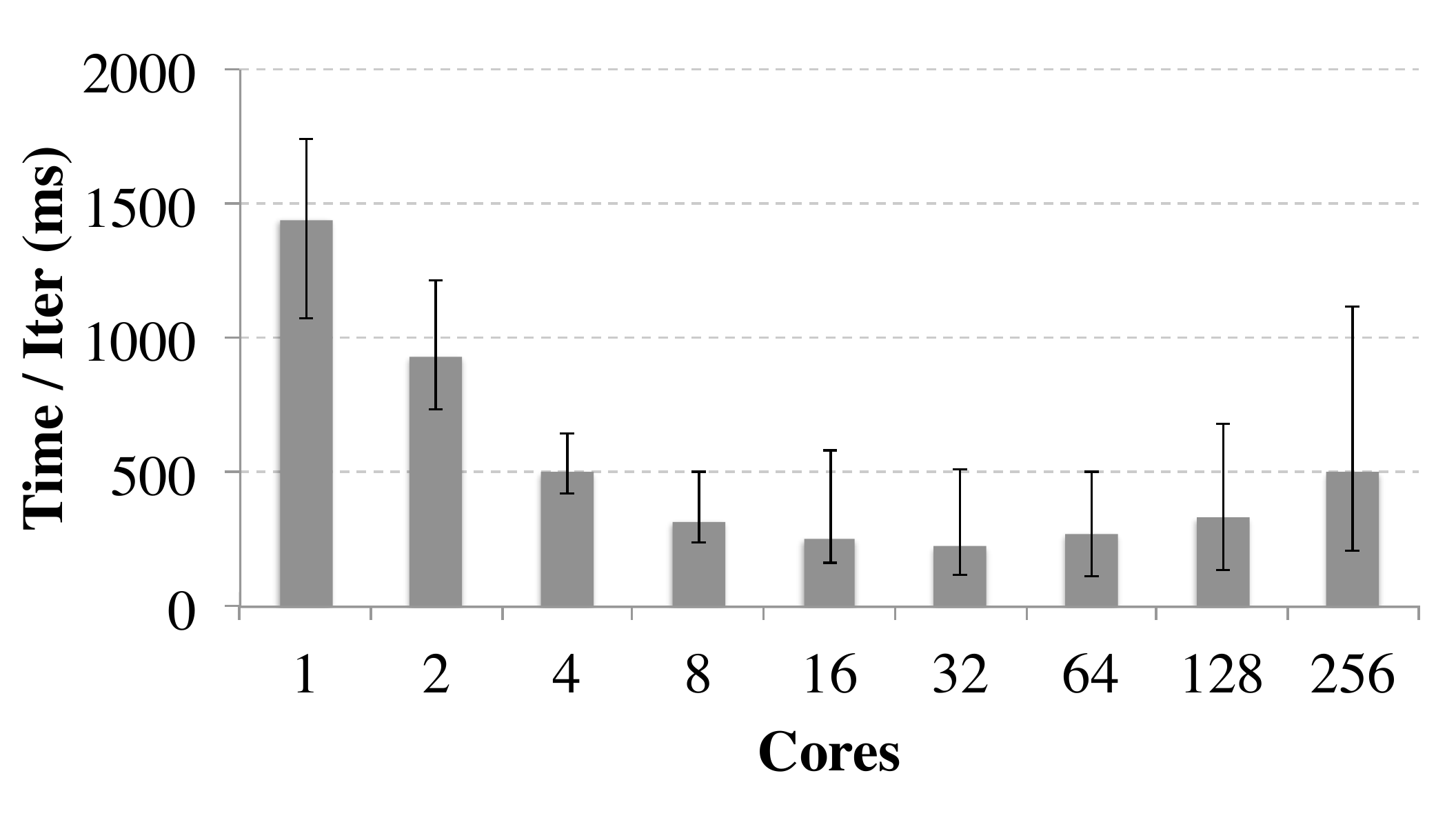}}
  \hspace{0.1in}
  \subfigure[Convergence of CoCoA as we vary the degree of parallelism used.]{\label{fig:convergence-parts}\includegraphics[width=0.30\textwidth]{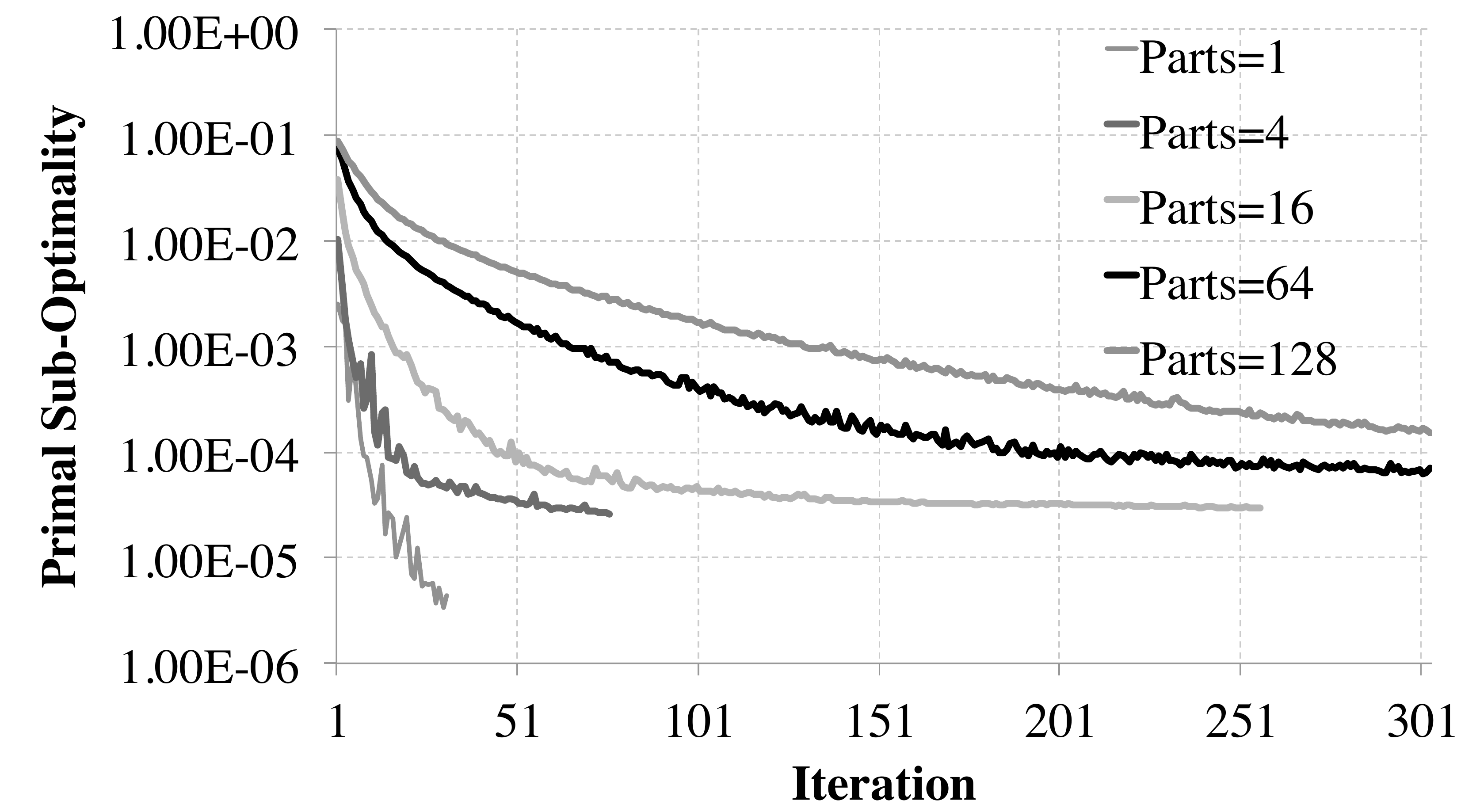}}
  \hspace{0.1in}
  \subfigure[Comparison of convergence rate of CoCoA, CoCoA+, SGD and Splash when using 16 cores.]{\label{fig:cocoa-sgd}\includegraphics[width=0.30\textwidth]{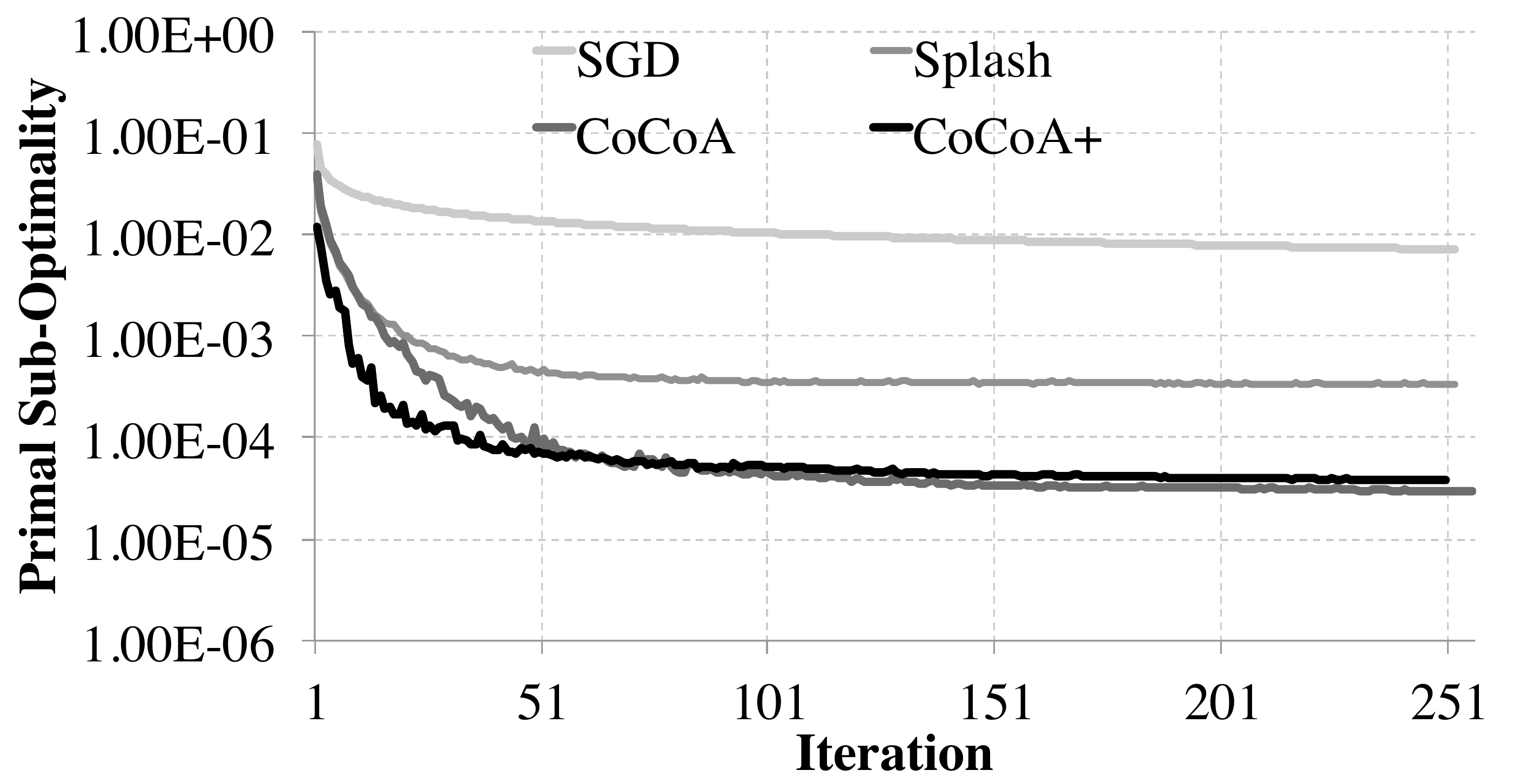}}
  \caption{Performance and convergence experiments using CoCoA.}
  \label{fig:motivation}
\end{figure*}

\subsection{Distributed computing}
The widespread adoption of cloud computing platforms like Amazon EC2, Microsoft Azure, Google
Compute Engine etc., means that users can now choose their computing substrate in terms 
of the number of cores, memory per machine and also in terms of the number of machines to use. However having
additional choice comes with its own challenges; the performance of machine learning jobs can vary
significantly based on the resources chosen and thus a number of recent
efforts~\cite{herodotou2011no, venkataraman2016ernest} have focused on recommending the best cluster configuration for
a given workload.

One of the important decisions users have to make is choosing the cluster size or the degree of parallelism
to use. Having a higher degree of parallelism usually lowers the \emph{computation time} but could
increase the amount of time spent in \emph{communication}. Specifically in the context of iterative
optimization algorithms, varying the degree of parallelism changes the time taken per-iteration and
we study its impact in Section~\ref{sec:case-study}.

\subsection{Distributed optimization algorithms}
Large scale optimization algorithms used in practice include first-order methods based on parallel
SGD~\cite{zhang15b, bottou10, zinkevich11}, coordinate descent methods~\cite{jaggi14,ma15,yang13} and
quasi-newton methods like L-BFGS~\cite{mokhtari14, moritz16}. These algorithms are typically
iterative and each iteration can be expressed using a bulk-synchronous step in a distributed framework
like Hadoop MapReduce or Spark~\cite{zaharia12}.

One of the main differences among the various algorithms is how their convergence rates change as
the degree of parallelism increases. For, methods like full-gradient descent (GD) where the gradient is
evaluated on all the data points at each iteration, the convergence rate remains the same
irrespective of the parallelism. However this is not the case for stochastic methods like
mini-batch SGD. For mini-batch SGD with batch size $b$, the optimization error after running 
for $T$ iterations is $O(1/\sqrt{bT} + 1/T)$~\cite{li2014efficient, dekel2012optimal}. 
Thus although $b$ times more examples are processed in an iteration, the mini-batch method
only achieves a $O(\sqrt{b})$ improvement in convergence when compared to serial SGD where one example
is processed at a time. Thus as the mini-batch size is increased in large clusters, the convergence 
rate, in terms of number of examples examined, typically degrades.

Similar effects can be seen in other algorithms. Recent work in CoCoA~\cite{jaggi14}, CoCoA+~\cite{ma15} perform local
updates using coordinate descent~\cite{shalevshwartz13} and obtain convergence rates that only
degrade with the number of machines rather than the mini-batch size. This is advantageous as
the number of machines is usually smaller than the mini-batch size which is order of data size.
Similar rates have also been shown for re-weighted stochastic gradient methods~\cite{zhang15b}.
In summary we see that for stochastic methods while increasing the degree of parallelism can
improve performance, the convergence rates degrade and thus users need to make a careful trade-off
in choosing the appropriate configuration.

Finally, the rates discussed in the previous paragraph are upper bounds and are usually
applicable for the worst-case inputs. However in practice, input data can behave much better and it
is difficult for users to accurately predict how each algorithm will
perform for a new dataset.

\vspace{-0.3cm}
\subsection{Case Study}
\label{sec:case-study}
To highlight the convergence and performance variations, we perform a simple binary classification
experiment to predict a single digit ($5$) using the MNIST dataset.
We ran CoCoA~\cite{jaggi14} on Apache Spark~\cite{zaharia12} with 
linear SVM as the loss function. We measure performance in terms of time per outer iteration of CoCoA and
also compute the primal sub-optimality at the end of every iteration. The experiments were run on a
eight-node YARN cluster where each machine has 48 cores, 256 GB RAM and 480 GB SSD storage. The machines
are further partitioned into Spark executors that each have 4 cores and 20GB of RAM each. We vary the
degree of parallelism by changing the number of executors used.
We run the algorithm until the primal sub-optimality reaches $1e-4$ or 500 iterations are completed and results from varying the
degree of parallelism are shown in Figure~\ref{fig:motivation}.

From Figure~\ref{fig:time-per-iter}, we can see that while the time taken per iteration goes down as
we scale from $1$ to $32$, but that performance degrades as we use more than $32$ cores. Further,
even in the regime where performance improves, we see that improvements are not linear; i.e.
doubling the number of cores does not result in halving the time per iteration. Both of these
effects are due to the computation vs. communication balance in the system. For MNIST, a small
dataset with just 60000 rows, we see that communication costs start to dominate with higher number
of cores and this leads to the poor scaling behavior. Similar effects have been observed in prior
work for larger datasets as well~\cite{venkataraman2016ernest}.

In Figure~\ref{fig:convergence-parts} we see that how the primal sub-optimality across iterations
changes as we vary the degree of parallelism. In this case we see that using $1$ core means that the
algorithm converges in around 10 iterations while using $16$ cores takes around $50$ iterations to
converge to $1E-4$ sub-optimality. This shows how the convergence rate can degrade as we increase
the degree of parallelism and why it is important for users to tune this appropriately.

Finally, as known in theory, the observed convergence rate in practice also varies depending
on the specific algorithm used. Figure~\ref{fig:cocoa-sgd} shows the convergence
for CoCoA, CoCoA+, Splash~\cite{zhang15b} and parallel SGD with local updates while using 16 cores. From the figure we
see that both CoCoA and CoCoA+ perform much better than SGD-based methods. We also see that while CoCoA+ converges
faster in the first 50 iterations, CoCoA performs slightly better beyond 50 iterations. Thus based on
the desired level of convergence and data properties the appropriate algorithm to use can
differ.


\section{Modeling Optimization Algorithms}
\label{sec:design}

In the previous section we saw how the changing degree of parallelism affects performance and
convergence. We next describe our approach to addressing this problem in Hemingway. We begin 
with high level goals for our system and then discuss how we can break down the problem into two
parts: modeling the system and modeling the algorithm. 

\subsection{Goals and assumptions}
\begin{figure*}[t!]
  \centering
    \includegraphics[width=\textwidth]{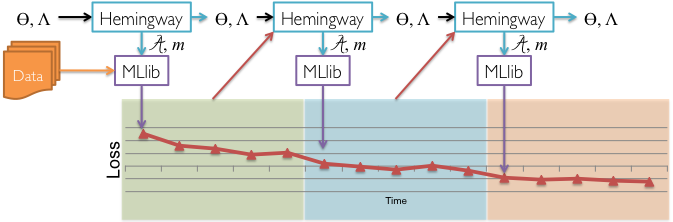}
    \caption{\footnotesize Idealized Example of using Hemingway.
             For each time frame, Hemingway takes as input the current estimated system model $\Theta$ and convergence model $\Lambda$, and suggests the best algorithm $\mathcal{A}$ and number of machines $m$ to use for the next time frame.
             These are then fed into a machine learning framework, e.g. MLlib, which executes the convex optimation algorithm.
             Resultant losses are provided as input into Hemingway to update $\Theta$ and $\Lambda$.
             }
    \label{fig:hemingway_ideal}
  \vspace{-0.1in}
\end{figure*}

The goal of our system, Hemingway, is to develop an interface where users can specify an end-to-end
requirements and the system will automatically select the appropriate algorithm and degree of parallelism to use. Such a
system can be used along with existing libraries of machine learning algorithms like
MLlib~\cite{meng2016mllib}, Vowpal Wabbit~\cite{langford2007vowpal}, SystemML~\cite{ghoting2011systemml} etc.
Specifically we would like to support use cases of the form: given a relative error goal of
$\epsilon$, choose the fastest algorithm and configuration; or given a target latency of $t$ seconds
choose an algorithm that will achieve the minimum training loss. An idealized example of our system
is shown in Figure~\ref{fig:hemingway_ideal}.

Our assumption is that the dataset used by the distributed optimization algorithm is
static and that the algorithm is iterative with each iteration expressed as a bulk synchronous paralllel (BSP) \cite{valiant1990bridging} job\footnote{
In a BSP job, machines iteratively perform local computations prior to communicating information (typically an updated model or gradients) with other machines.
Crucially, a synchronization barrier ensures all machines complete their computations and communications for the iteration before the next iteration begins}.
We also assume homogeneous machines and network\footnote{In practice, we can approximate this well using the same AWS instance type.}.
Further our current study is restricted to convex loss functions.
We discuss extensions to these assumptions in Section~\ref{sec:challenges}.

\subsection{Overall model}
We denote the objective value attained by an algorithm after running for time $t$ when running on $m$ machines
as $h(t, m)$. Our goal is to build a model that can predict the value of $h$ so that we can compare different
configurations and optimization algorithms. Our key insight in this work is that we can split this task into
a decomposition of two models: a system-level model $f(m)$ that returns the time taken per iteration given $m$ machines
and a model for convergence $g(i, m)$ that predicts the objective value after $i$ iterations given $m$ machines.
Thus the overall model can be obtained by combining our above two models: $h(t, m) = g(t / f(m), m)$.

The main benefit of this approach is that we can train the two models independently and reuse them based on
changes. For example if there are new machine types or networking hardware changes in a datacenter we can retrain 
just the system model and reuse the convergence model. Similarly if there are changes to the algorithm in
terms of parameter tuning, we only retrain the convergence model.

\subsubsection{Modeling the system}
To build the system level model, we propose re-using the approach in
Ernest~\cite{venkataraman2016ernest}, a performance prediction framework that minimizes the time and
resources spent in building a performance model. We summarize the main ideas in Ernest below and
specifically on how it applies to ML algorithms.

The response variable in our model is the running time $f(m)$ of each BSP iteration as a function of number of machines $m$.
The functional form of $f$ is guided by different components and how they influence the time taken
for an iteration. The components include the time spent in computation, which scales inversely to the
number of machines and time taken by common communication patterns like broadcast, tree-reduction
and shuffles, which increase as we increase the number of machines. Putting these terms together the
model for an algorithm like mini-batch SGD or CoCoA looks like
\[
f(m)  = \theta_{0} + \theta_{1} \times (size / m) + \theta_{2} \times \log(m) + \theta_{3} \times m
\]
where $size$ represents the amount of input data and $m$ represents the number of machines used.
Ernest fits the above model by measuring performance on small samples of data and using the model we
can then extrapolate the time taken at larger scales. Results in~\cite{venkataraman2016ernest} show
that the prediction error for the time taken per-iteration of mini-batch SGD is within 12\%, 
while using samples smaller than 10\% of input data.

For certain ML algorithms, we may need to modify the above model to include terms that reflect other
computation or communication patterns. For example while the computation costs in first order 
methods typically scale linearly with number of examples, using second order methods 
like SDNA~\cite{qu2015sdna} could incur super-linear computation or communication costs.

\subsubsection{Modeling algorithm convergence}
Most popular optimization algorithms today take an iterative approach towards minimizing an objective value.
We capture this behavior with a bivariate function $g(i, m)$ that returns the objective value\footnote{
  We are usually interested in the objective value, but the Hemingway approach is data-driven and thus sufficiently flexible to handle other metrics such as test classification accuracy, precision, recall, etc.
}
after the algorithm is executed for $i$ iterations on $m$ machines.
Optimization algorithms are typically accompanied by analyses of upper bound convergence rates; these rates can help guide us in putting together a functional form for $g(i,m)$.
For example, CoCoA has a upper bound convergence rate of
$g(i,m) \leq \left(1 - \frac{c_0}{m} \right)^i c_1$,
where $c_0$ and $c_1$ are data-dependent constants.
We point out, however, that the actual observed convergence rates can differ from the theoretical upper bounds, so it is important not to overly constrain $g$'s functional form.
In this paper, we have assumed linear forms of $g$ for ease of fitting with ordinary least squares or Lasso.
For ease of fitting with ordinary least squares or Lasso, we assume in this paper a linear form for $g = \sum_{j=1}^k \lambda_j \phi_j(i,m)$, where $\lambda_j$'s are parameters to be estimated, and $\phi_j(i,m)$'s are possibly non-linear features.
Non-linear functional forms of $g$ can, in general, be fitted by minimizing the squared error
between the model and the data.


\section{Preliminary Experiments With CoCoA+}

We demonstrate the ability of Hemingway to model algorithm convergence by fitting a linear model to an example run of CoCoA+.
We used CoCoA+ to solve a binary classification problem on MNIST, and varied the degree of parallelism $m$ from 1 to 128 in powers of 2.
The algorithm was terminated when the primal sub-optimality reached $1e-4$, or after 500 iterations.
We then fit a linear model to $\log(P(i,m) - P^*)$ using using \texttt{LassoCV} from scikit-learn \cite{scikit-learn}, where $P(i,m)$ is the primal objective value at iteration $i$ with $m$ parallelism.
A range of fractional, polynomial, and logarithmic terms were used as the features of our model.

We also collected data for estimating the Ernest system model using Amazon EC2 R3.xlarge instances with 30.5GB RAM and 4 cores each.

Figure\footnote{\label{fn:ref100iters}Plots showing the first 100 iterations are provided in Appendix \ref{app:experiments}.} \ref{fig:pred-train-alliters} shows the fit of our learned Hemingway model, and illustrates that we are able to capture the convergence trends exhibited by CoCoA+.
We were also able to combine the Ernest and Hemingway models together to capture the convergence trends as a function of time, as seen in Figure \ref{fig:pred-train-alltimes}.

For the Hemingway model to be practically useful, we need to be able to use it for predicting convergence at unobserved settings of $m$ and $i$.
We consider two such scenarios below.

\begin{figure*}[!t]
  \centering
    \subfigure[Hemingway prediction across iterations.]{\label{fig:pred-train-alliters}\includegraphics[width=0.48\textwidth]{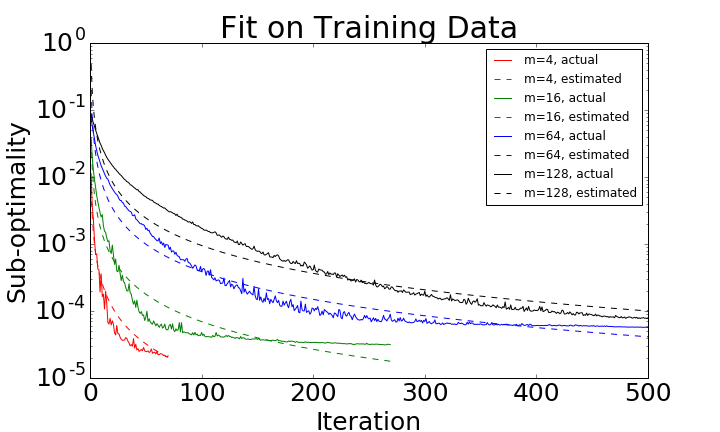}}
    \subfigure[Ernest Hemingway prediction across time.]{\label{fig:pred-train-alltimes}\includegraphics[width=0.48\textwidth]{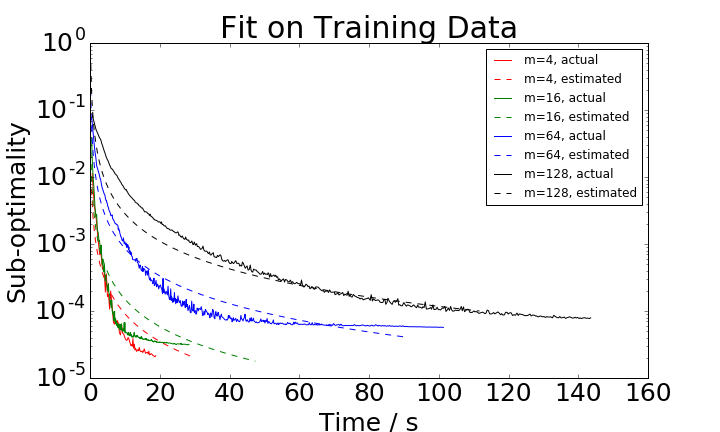}}
    \caption{\footnotesize Convergence of CoCoA+ compared against our fitted model.}
  \vspace{-0.2in}
\end{figure*}

\subsection{Predicting for unobserved degree of parallelism}
\begin{figure*}[!t]
  \centering
  \subfigure[Hemingway prediction across iterations.]{\label{fig:pred-loocv-alliters}\includegraphics[width=0.48\textwidth]{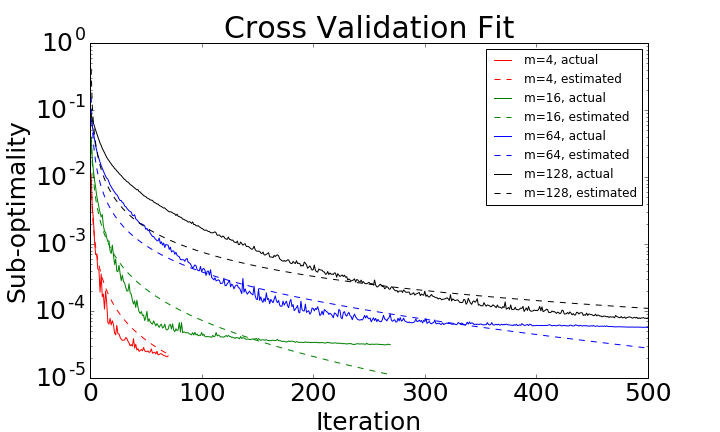}}
  \subfigure[Ernest Hemingway prediction across time.]{\label{fig:pred-loocv-alltimes}\includegraphics[width=0.48\textwidth]{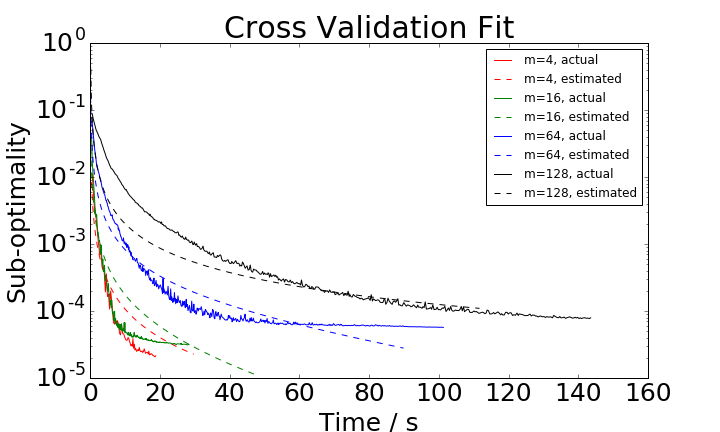}}
  \caption{\footnotesize Convergence of CoCoA+ compared against our fitted model when predicting for unobserved degree of parallelism.}
  \vspace{-0.2in}
\end{figure*}

In the first scenario, we have collected convergence results for some values of $m$, and would like to predict the convergence trend for a yet-unobserved degree of parallelism.
This can be simulated by a leave-one-$m$-out cross validation with our data.
For example, we predict the convergence $g(i, 128)$ for $m=128$ parallelism by using data collected from $m=1,2,4,8,16,32,64$.
Figure\cref{fn:ref100iters} \ref{fig:pred-loocv-alliters} shows the resultant cross validated models are good fits for the true convergence.
Hence, we are able to estimate the trend of convergence for unobserved values of $m$.

\subsection{Forward prediction}
\begin{figure*}[!t]
  \centering
  \subfigure[]{\label{fig:pred-next01-alliters}\includegraphics[width=0.48\textwidth]{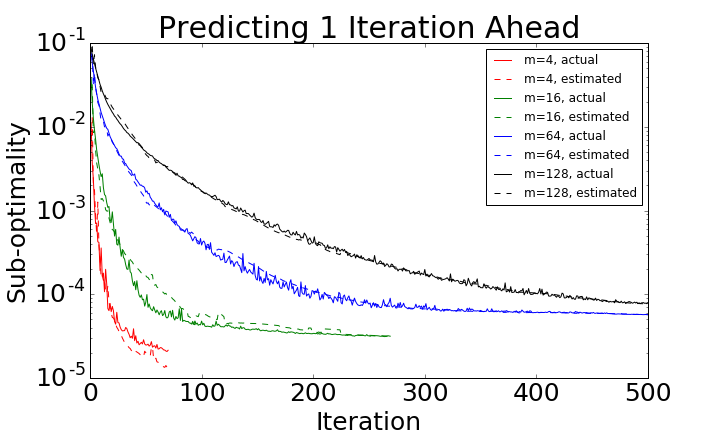}}
  \subfigure[]{\label{fig:pred-next10-alliters}\includegraphics[width=0.48\textwidth]{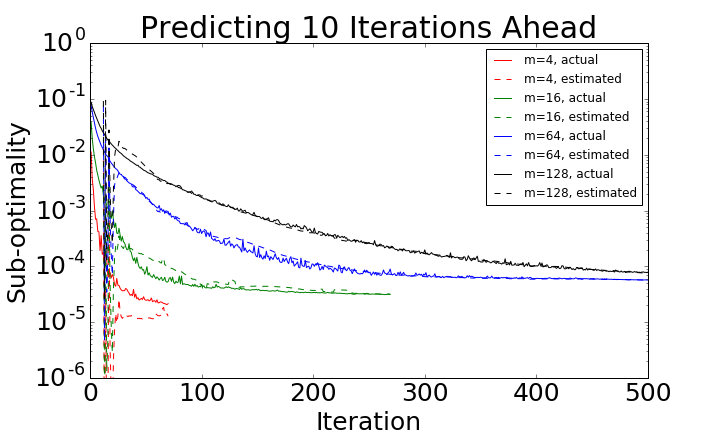}}
  \caption{\footnotesize Convergence of CoCoA+ compared against our fitted model when predicting for future iterations.}
  \vspace{-0.2in}
\end{figure*}

Secondly, we consider prediction of convergence at future iterations.
Figures\cref{fn:ref100iters} \ref{fig:pred-next01-alliters} and \ref{fig:pred-next10-alliters} respectively show the fit of models for predicting 1 and 10 iterations ahead, given a window of 50 iterations in the past.
In both cases, predictions become more accurate when $i$ is sufficiently large to provide enough information for modeling.

\begin{figure*}[!t]
  \centering
  \subfigure[]{\label{fig:pred-next1000ms-alltimes}\includegraphics[width=0.48\textwidth]{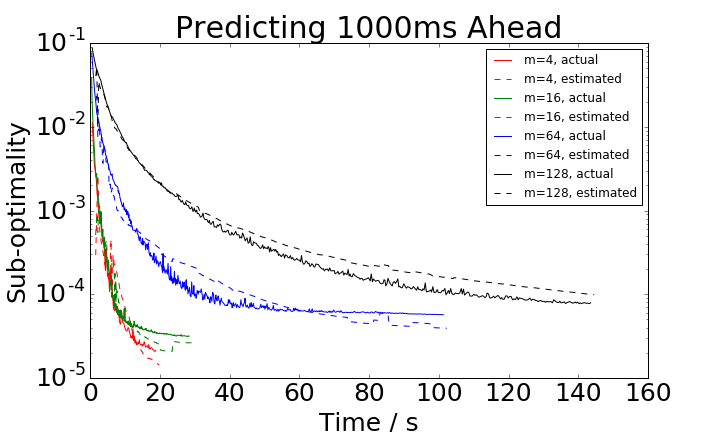}}
  \subfigure[]{\label{fig:pred-next5000ms-alltimes}\includegraphics[width=0.48\textwidth]{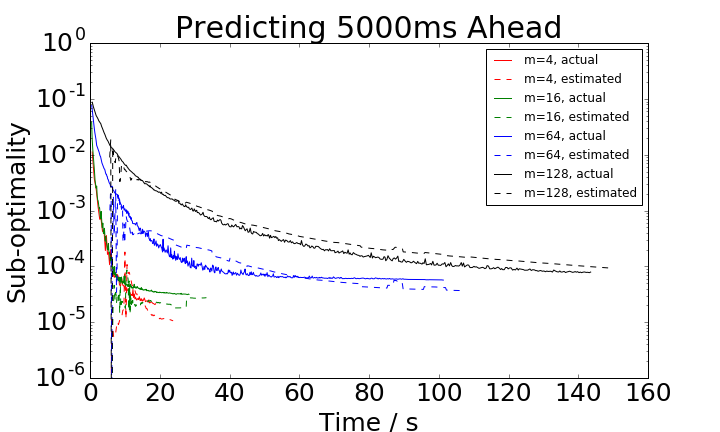}}
  \caption{\footnotesize Convergence of CoCoA+ compared against our fitted model when predicting in future time.}
  \vspace{-0.2in}
\end{figure*}

We also applied the Ernest and Hemingway models in combination to predict convergence in future time.
Figures\cref{fn:ref100iters} \ref{fig:pred-next1000ms-alltimes} and \ref{fig:pred-next5000ms-alltimes} show that our models can capture the convergence trends at 1s and 5s in the future.

\section{Related Work}

Similar decoupling of systems and algorithmic efficiencies had been proposed in DimmWitted~\cite{zhang2014dimmwitted} for studying running times of optimization algorithms in NUMA machines.
Tradeoffs of different access methods, and data and model replication techniques were examined in
terms of their statistical and hardware efficiencies. 
In Hemingway, we similarly propose decoupling the statistical and hardware execution models, but 
additionally we also propose an automatic method to choose the best configuration given a new
algorithm.

A number of previous studies have looked at optimizing the performance of large scale
machine learning algorithms. On the systems side, physical execution engines that can 
choose between row or column storage layouts~\cite{zhang2014dimmwitted}, and choose between
algorithms given a fixed iteration budget~\cite{sparks2016keystone} have been proposed. 
Further systems like STRADS~\cite{kim2016strads} also study the trade-offs between using 
model parallelism and data parallelism with parameter servers.
In Hemingway we focus on black-box modeling convergence rates of distributed optimization 
algorithms as opposed to physical execution and our formulation can reuse 
the execution models developed by existing systems. 

Recently, distributed deep learning frameworks like Tensorflow~\cite{tensorflow2015-whitepaper} and
MXNET~\cite{chen2015mxnet} have implemented various execution strategies~\cite{miliagkas2016async}. As a consequence,
systems like \cite{paleo} have been developed to model the cost of each operator and to choose the optimal
number of CPUs, GPUs~\cite{hadjis2016omnivore} given asynchronous execution. 
These systems are typically ``white-box,''
i.e. they exploit the characteristics of specific algorithms like asynchronous SGD and specific operators in a 
convolutional neural network. In Hemingway we proposed using a black-box model that can be applied across a number
of distributed optimization algorithms.

There has been renewed interest in the machine learning community in ``meta-learning,'' where machine learning techniques are used to automatically learn optimal models
\cite{andreas2016learning,baker2016designing}
or optimization algorithms
\cite{andrychowicz2016learning,duan2016rl,chen2016learning}.
In particular, \cite{chen2016learning} attempts to learn an optimizer for black-box functions.
Meta-learning approaches typically require much more data and training than Hemingway, since the meta-learner models are usually deep neural networks with large numbers of parameters.
Furthermore, the objective of these meta-learners is to minimize the underlying objective function, and usually do not model or account for the system efficiency.


\section{Challenges}
\label{sec:challenges}
While our initial experiments show promise in terms of the utility and insights that can be gained
from modeling convergence rates, there are number of challenges we are addressing to make this
system practical.

\textbf{Training time.} One of the important aspects of any modeling based approach is the amount
of time it takes to train the model before it can be make useful suggestions to the user. While our
initial experiments have shown that convergence rates can be extrapolated across iterations and
partition sizes, we plan to study if techniques to minimize data acquisition like
experiment design can be used to minimize the time spent in data collection.

\textbf{Training resources.} Closely related to the time it takes to train a model, is the amount of
resources used to train a model. For cloud computing settings, this is especially important as
launching a large number of machines to collect training data could be expensive. While prior work
in Ernest~\cite{venkataraman2016ernest} discussed how the system-level model can be trained using a
small number of machines and data samples, we plan to investigate if similar approximations can be
made for the convergence model. This would similar to bootstrap where we would try to extrapolate
the convergence model on the entire dataset based on the rates observed on a random subset of the
data.

\textbf{Adaptive algorithms.} We believe that using our modeling approach we can also study
the development on new algorithms that adapt based on the requirements. For example while using a
large number of cores might be appropriate at the beginning while the function value is far from
optimal, we can then adaptively change the degree of parallelism as we get closer to convergence.
Decisions on whether it is worthwhile to make such a change and when such changes should be made can
be taken using the models we build.

\textbf{Asynchronous algorithms.} While BSP algorithms have clear synchronization barriers between
iterations, the same is not true of asynchronous algorithms such as Hogwild!
\cite{recht2011hogwild}.  Nevertheless, many of these algorithms have a natural notion of an
\textit{epoch}, typically comprised of a single pass over the entire dataset.  By using an epoch as
a unit of work done, we believe it is still possible to model both the algorithm convergence and
system performance.  Further investigation is required to see if linear models will suffice,
or if more complex modeling, e.g. using queueing theory, is required.

\textbf{Non-Convex loss functions.} Finally while our current efforts are focused on convex loss
functions like logistic or hinge, we also plan to study if similar ideas can be used to model
optimization of non-convex functions used in settings like deep learning.

\section{Conclusion}
In this paper we studied how distributed optimization algorithms scale in terms of
performance and convergence rate and showed how choosing the optimal configuration
could significantly impact training time. To address this, we propose Hemingway, a system
that models the convergence rate for distributed optimization algorithms and we present some 
of the challenges involved in building such a system.

{
\bibliographystyle{abbrv} 
\bibliography{main}
}

\newpage
\appendix
\section{Additional Plots For CoCoA+ Experiments}
\label{app:experiments}

\begin{figure*}[ht]
  \centering
    \subfigure[Hemingway prediction across iterations.]{\label{fig:pred-train-iters-100}\includegraphics[width=0.48\textwidth]{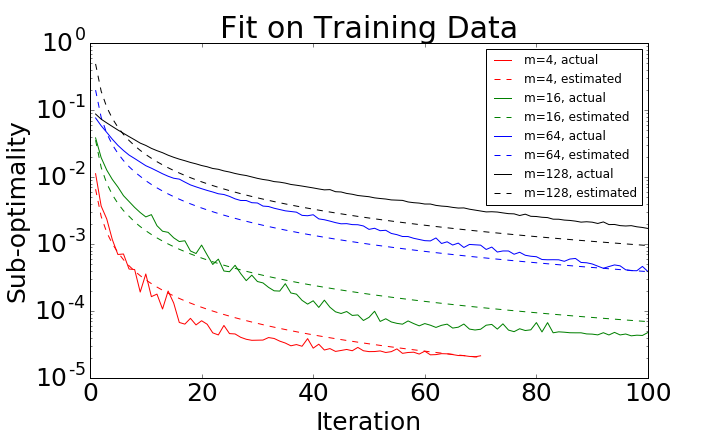}}
    \subfigure[Ernest Hemingway prediction across time.]{\label{fig:pred-train-times-100}\includegraphics[width=0.48\textwidth]{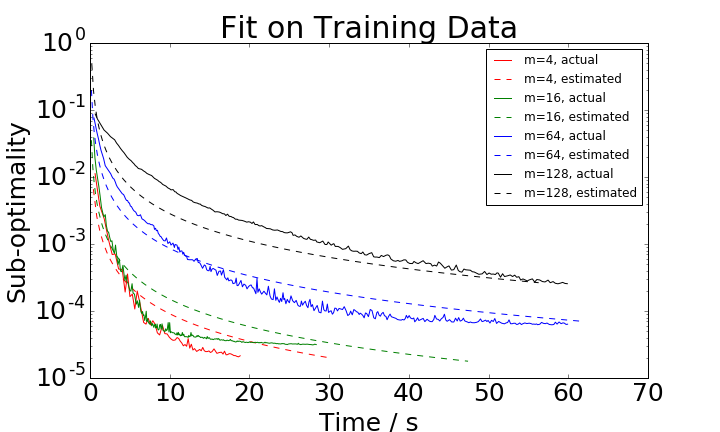}}
  \caption{Convergence of CoCoA+ compared against our fitted model.}
\end{figure*}

\begin{figure*}[ht]
  \centering
    \subfigure[Hemingway prediction across iterations.]{\label{fig:pred-loocv-iters-100}\includegraphics[width=0.48\textwidth]{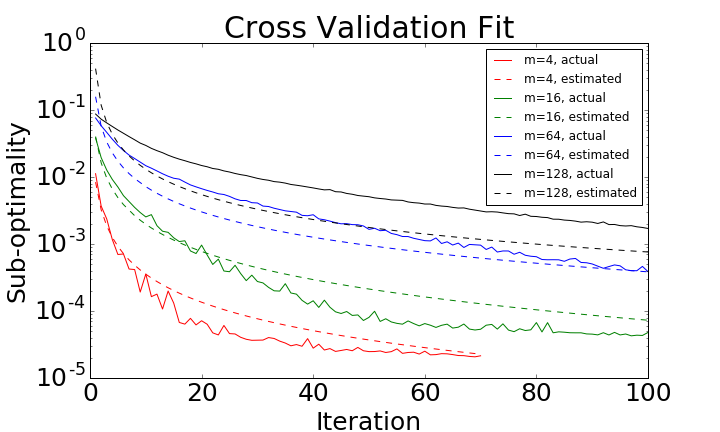}}
    \subfigure[Ernest Hemingway prediction across time.]{\label{fig:pred-loocv-times-100}\includegraphics[width=0.48\textwidth]{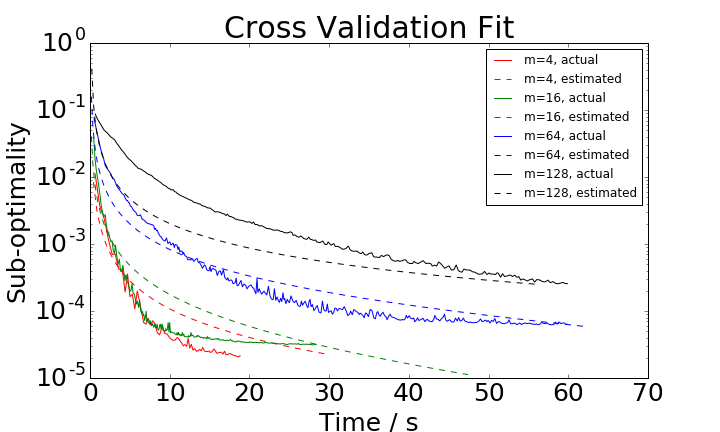}}
  \caption{Convergence of CoCoA+ compared against our models for unobserved degrees of parallelism.}
\end{figure*}

\begin{figure*}[ht]
  \centering
  \subfigure[]{\label{fig:pred-next01-100}\includegraphics[width=0.49\textwidth]{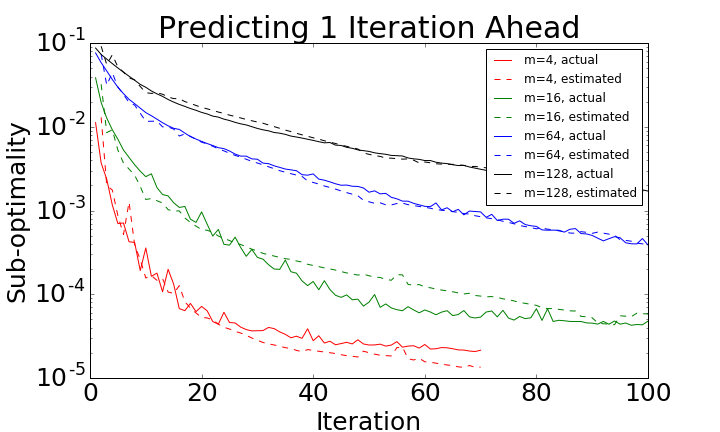}}
  \subfigure[]{\label{fig:pred-next10-100}\includegraphics[width=0.49\textwidth]{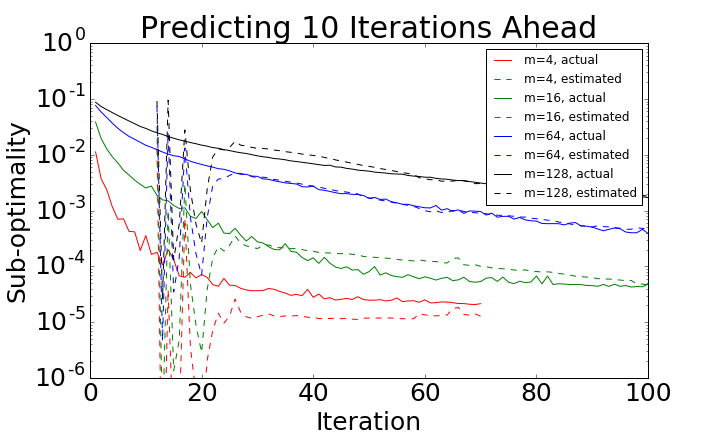}}
  \caption{Convergence of CoCoA+ compared against our forward-prediction models.}
\end{figure*}

\begin{figure*}[ht]
  \centering
  \subfigure[]{\label{fig:pred-next1000ms-1m}\includegraphics[width=0.49\textwidth]{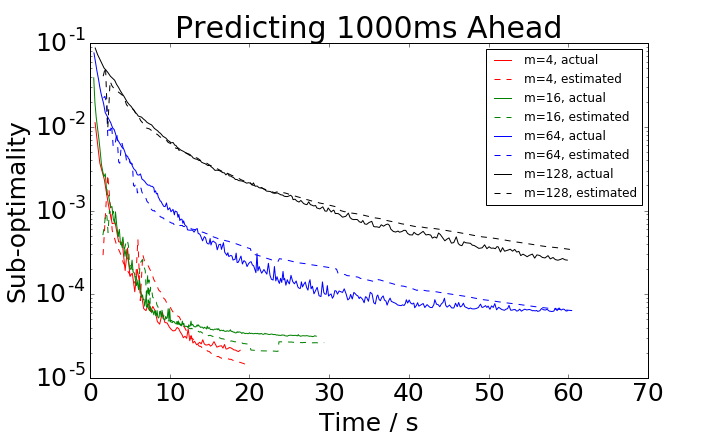}}
  \subfigure[]{\label{fig:pred-next5000ms-1m}\includegraphics[width=0.49\textwidth]{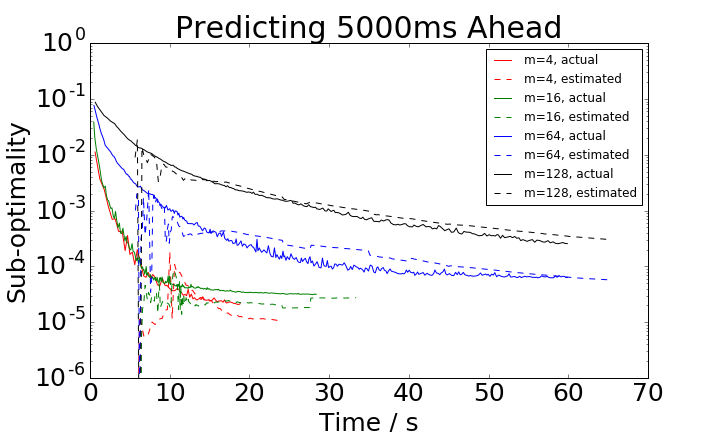}}
  \caption{Convergence of CoCoA+ compared against our forward-prediction models.}
\end{figure*}

\end{document}